\documentclass[11pt,a4paper]{article}
\usepackage[margin=3.5cm]{geometry}

\usepackage{amsmath,amssymb,cancel}
\usepackage{graphicx,hyperref}
\usepackage{epstopdf}

\usepackage{authblk}

\begin{document}
	\title{Renormalization on noncommutative torus}
	
	
	\author[a]{D. D'Ascanio\thanks{\tiny dascanio@fisica.unlp.edu.ar}}
	\author[a]{P. Pisani\thanks{\tiny pisani@fisica.unlp.edu.ar}}
	\author[b,c]{and D. V. Vassilevich\thanks{\tiny dvassil@gmail.com}}
	
	\affil[a]{\small Instituto de F\'isica La Plata, CONICET -- Universidad Nacional de La Plata,
		
		C.C. 67 (1900), La Plata, Argentina}
	\affil[b]{\small CMCC, Universidade Federal do ABC,
		CEP 09210-180, Santo Andr\'e, S.P., Brazil}
	\affil[c]{\small Department of Physics, Tomsk State University, Tomsk, Russia}
	
	
	
	
	
	\maketitle
	
	\abstract{We study a self-interacting scalar $\varphi^4$ theory on the $d$-dimensional noncommutative torus. We determine, for the particular cases $d=2$ and $d=4$, the counterterms required by one-loop renormalization. We discuss higher loops in two dimensions and two-loop contributions to the self-energy in four dimensions. Our analysis points toward the absence of any problems related to the UV/IR mixing and thus to renormalizability of the theory. However, we find another potentially troubling phenomenon which is a wild behavior of the two-point amplitude as a function of the noncommutativity matrix $\theta$.}


\section{Introduction}
One of the motivations for considering quantum field theories on noncommutative spaces was the hope that they may be ultraviolet (UV) finite.
It was shown, however, that UV divergences persist on the noncommutative (NC) Moyal plane \cite{Filk:1996dm,Chaichian:1998kp}. Moreover, though
certain Feynman diagrams are less UV divergent than in the commutative case,  they develop
singularities at some special, typically zero, value of the external momenta. When such diagrams
appear as subgraphs of higher-order diagrams, the latter diagrams become divergent in a nonrenormalizable
manner. This phenomenon \cite{Chepelev:1999tt,Minwalla:1999px,Aref'eva:1999sn}, called the UV/IR mixing
\cite{Minwalla:1999px}, is the main obstacle to renormalization of NC field theories.

It was believed for some time that the UV/IR mixing appears exclusively in Euclidean signature spaces.
However, it was demonstrated \cite{Bahns:2010dx} that similar problems exist in Minkowski
spacetime as well.

Various methods were proposed to deal with this problem. Of course, the supersymmetry helps to achieve renormalizability of noncommutative theories \cite{SheikhJabbari:1999iw,Girotti:2000gc}. Grosse and Wulkenhaar \cite{Grosse:2003nw,Grosse:2004yu} motivated by the Langmann-Szabo duality \cite{Langmann:2002cc} proposed to add to the action an oscillator term which breaks translation invariance but ensures renormalizability. Modifications of the momentum dependence of the kinetic term were considered in \cite{Gurau:2008vd}. Taking the noncommutativity parameter nilpotent \cite{Fresneda:2008sr} also improves renormalization. It has been shown \cite{Ruiz} that spontaneous symmetry breaking softens the UV/IR mixing. A fairly recent review is Ref.\ \cite{Blaschke:2010kw}.

In this work, we take a different path. We consider a noncommutative $\varphi^4$ theory on a torus. Sensitivity of UV divergences in NC theories to the presence of compact dimension (and even eventual disappearance of such divergences) has been stressed already in \cite{Chaichian:1998kp}; see also \cite{Chaichian:2001pw}. Note, however, that, due to a different implementation of noncommutativity, the existence of a compact dimension in the two-dimensional case considered in \cite{Chaichian:1998kp} guarantees the finiteness of tadpole contributions. This is not the case in the model considered in the present article, where quantum corrections are UV-divergent and must be properly renormalized.

One may get an idea on the structure of counterterms, singularities of the propagators etc. by looking at the heat kernel expansion (see, e.g. \cite{Vassilevich:2003xt}). Roughly speaking, the relevant operators\footnote{For bosonic theories, these are the operators $L$ appearing in the second variation of classical action, $S_2=\int (\delta \varphi) L(\varphi) (\delta\varphi)$, with $\delta\varphi$ being a fluctuation and $\varphi$ -- a background field.} on noncommutative spaces are generalized Laplacians that contain gauge fields and potentials (as usual Laplacians), but these gauge fields and potentials act by left or right Moyal multiplications on the fluctuations $\delta\varphi$. If the generalized Laplacian contains only left or only right Moyal multiplications, the structure of corresponding heat kernel coefficients is very simple on both NC torus \cite{Vassilevich:2003yz} and NC plane \cite{Gayral:2004ww} -- they look almost as their commutative counterparts with star products instead of usual products. For interesting theories, however, the relevant operators contain \emph{both} left and right multiplications. For such operators on the NC plane the structure of heat kernel coefficients is very complicated
\cite{Vassilevich:2005vk,Bonezzi:2012vr} thus reflecting the presence of the UV/IR mixing. The situation changes drastically on NC torus \cite{Gayral:2006vd}. If the noncommutativity parameter satisfies the so-called Diophantine condition or is rational, the heat kernel coefficients (and thus the one-loop counterterms) assume a very simple form if written in terms of a suitably defined trace operation on the algebra of smooth functions on the torus. We shall use this observation to formulate our proposal for a (presumably) renormalizable $\varphi^4$ theory on NC torus.

Let us stress that the notion of locality does not make much sense in noncommutative theories since the star product itself is nonlocal. Instead of local polynomial actions one has to use traces of polynomials constructed from the fields and their derivatives. There are more different traces on $\mathbb{T}^d_\theta$ than on $\mathbb{R}_\theta^d$. This observation will be crucial for our construction of admissible counterterms.

Here we like to mention several papers that considered quantum field theories on NC torus. In Ref. \cite{SheikhJabbari:1999iw} it was demonstrated that supersymmetric Yang-Mills theory on $\mathbb{T}^3_\theta$ with rational $\theta$ is one-loop renormalizable. Pure Yang-Mills theories were considered in \cite{Krajewski:1999ja} at one loop. Some arguments regarding the higher-order behavior were also presented. Relations between NC theories on $\mathbb{T}^d_\theta$ with rational $\theta$ and matrix models were studied in \cite{Guralnik:2002ru,Landi:2004sc}.

The purpose of this paper is to set up the stage for renormalization on NC torus and to discuss basic features of this process. First, we write down the model and introduce new counterterms for Diophantine and rational $\theta$. We analyze in detail two- and four-point functions at one loop. In $d=2$, the only superficially divergent diagrams are the one-loop two-point functions. We demonstrate that the insertion of these diagrams (together with counterterms) into internal lines of other diagrams does not lead to any divergences, so that there is no UV/IR mixing (at least in its classical formulation \cite{Minwalla:1999px}) on $\mathbb{T}^2_\theta$. In $d=4$, we analyze the two-loop self-energy diagrams. All our findings, though do not contain a complete proof, strongly suggest that the introduction of new  counterterms does make the $\varphi^4$ theory on NC torus in $d=2$ and $d=4$ renormalizable.

The counterterms depend in a very essential way on the number theory nature of $\theta$. But not only this, we show that also renormalized two-point functions (too) strongly depend on $\theta$. More precisely, we compare the two-point functions in $d=2$ for two close values $\theta_1$ and $\theta_2$ of the noncommutativity matrix, one being rational, and the other - irrational (Diophantine). We find that the typical variation of the two-point function is $\sim \ln ||\theta_1 -\theta_2||$. However, this does not necessarily mean that the theory has no prediction power. We discuss the implications and possible ways out in the Conclusions of the article.

The paper is organized as follows. The next section contains the definitions that will be used throughout the text. In Sect.\ \ref{sec:1l2p} we consider the two-point functions at one-loop order and analyze their renormalization and variation with $\theta$. Sect.\ \ref{four-point} is dedicated to four-point functions at one loop. Higher loops in $d=2$ are considered in Sect.\ \ref{sec-2d} and two-loop two-point functions in $d=4$ in Sect. \ref{sec:2l4D}. Our results are discussed in Sect.\ \ref{sec:con}. The behavior of some double sums is analyzed in Appendices \ref{ap_upp} and \ref{ap_tp}.
\section{The model}\label{sec:model}
As a base manifold we take the $d$-dimensional noncommutative (NC) torus $\mathbb{T}^d_\theta$ with unit radii; see \cite{CTn}.
The algebra $\mathcal{A}_\theta$ of smooth functions on $\mathbb{T}^d_\theta$ is formed by the
Fourier-type series
\begin{align}
    \varphi=\sum_{p\in \mathbb{Z}^d} \varphi_p\,U_p \,,\label{phisum}
\end{align}
where the Fourier coefficients $\varphi_p\in\mathbb{C}$ vanish at $|p|\to \infty$ faster than any power of $p$.
The unitaries $U_p$ satisfy
\begin{align}
    U_p\,U_q=e^{i\pi\, p\theta q}\, U_{p+q}\,, \label{Upq}
\end{align}
where $\theta$ is a constant and non-degenerate skew-symmetric $d\times d$ matrix. Expressions such as $p\theta q$ represent the quadratic form $\theta^{\mu\nu}p_\mu q_\nu$. One may think of $U_p$'s as of plane waves
$e^{ipx}$. Then the well-known Moyal product
\begin{align}
    (\varphi \star \psi) (x) = \exp \left( - i\pi\,\theta^{\mu\nu}\partial_\mu^x \partial_\nu^y\right)
    \varphi(x)\psi(y) \vert_{y=x} \label{Moyal}
\end{align}
reproduces \eqref{Upq},
\begin{align}
    e^{ipx}\star e^{iqx}=e^{i\pi\, p \theta q}\, e^{i(p+q)x} \,.
\end{align}
One should keep in mind, however, that the Moyal star product has to be understood as a \emph{formal} expansion
in the noncommutativity parameter, i.e., it is not convergent.

There is a trace on the algebra $C^\infty (\mathbb{T}_\theta^d)$ defined through
\begin{align}
\tau (\varphi) =\int \frac{d^dx}{(2\pi)^d} \ \varphi\,, \label{deftau}
\end{align}
which can be implemented in $\mathcal{A}_\theta$ by $\tau (U_p)=\delta_{p}$.\footnote{We use $\delta_p$ to denote 1 if $p=0$, and 0 otherwise.}

To proceed, we need some number theory preliminaries concerning the matrix elements of $\theta$. Let us define the set
\begin{align}\label{theset}
  \mathcal{Z}_\theta=\left\{q\in \mathbb{Z}^d/\  \theta q\in \mathbb{Z}^d\right\}\,.
\end{align}
Clearly, $\mathcal{Z}_\theta$ is a $\mathbb{Z}$-linear space whose dimension is the rank of the rational part of $\theta$. As we will see, field modes $\varphi_p$ with momentum $p\in\mathcal{Z}_\theta$ present a distinct renormalization behavior in the sense that they are affected differently by quantum corrections.

On the other hand, it has been demonstrated in \cite{Gayral:2006vd} that the heat kernel expansion and thus the one-loop divergences in a wide range of quantum field theories on the NC torus are well under control if the ``irrational'' part of $\theta$ satisfies a certain Diophantine condition; namely, there should be two positive constants, $C$ and $\beta$, such that
\begin{align}
\inf_{k\in \mathbb{Z}^d} |\theta q-k| \ge \frac {C}{|q|^{1+\beta}} \label{Dioph}
\end{align}
for all $q\in \mathbb{Z}^d\backslash \mathcal{Z}_\theta$. In the last section of this article, we will see that this Diophantine condition becomes crucial for the determination of the divergences of the double sums corresponding to some two-loop Feynman diagrams.

Our starting point is then the following action for a self-interacting scalar particle on the NC torus:
\begin{align}
    S[\varphi]=\tfrac 12\,\tau(\partial\varphi\,\partial\varphi) + \tfrac12\,m^2\,\tau(\varphi^2)
    +\lambda\,\tau(\varphi^4)\,.\label{action}
\end{align}
All products in \eqref{action} are in the noncommutative algebra, i.e., they are star-products. Since these will be the only products used in this work, we shall never write the symbol $\star$ explicitly.
The scalar field $\varphi$ undergoes a self-interaction given by the four-point vertex which in Fourier space can be written as
\begin{align}
    \lambda\,\sum_{k_1,\ldots,k_4}
    \delta_{k_1+\ldots+k_4}\,
    e^{i\pi(k_1\theta k_2+k_3\theta k_4)}\,
    \varphi_{k_1}\varphi_{k_2}\varphi_{k_3}\varphi_{k_4}\,.\label{lambda}
\end{align}
The free propagator is given by
\begin{align}\label{free-propagator}
    \langle\varphi_p\varphi_{p'}\rangle_{\rm free}=\frac{\delta_{p+p'}}{p^2+m^2}\,.
\end{align}

The heat kernel analysis of \cite{Gayral:2006vd} suggests that the theory may be renormalized by adding the counterterm action\footnote{These terms are certain (Dixmier-type) traces on the NC torus; see \cite{Gayral:2006vd}. In these sense, they generalize the trace terms in \eqref{action}. They may also be interpreted as usual traces after projecting the fields to a subalgebra \cite{Vassilevich:2007gt}.}
\begin{equation}
S_{\rm c.t.}= \sum_{p\in \mathcal{Z}_\theta} \left( \frac {\mu^2}2\, (\varphi)_p\, (\varphi)_{-p}
+ \lambda_1\, (\varphi)_p\, (\varphi^3)_{-p} + \lambda_2\, (\varphi^2)_p\, (\varphi^2)_{-p} \right)\,
\label{ctaction}
\end{equation}
(in addition to counterterms for the couplings in \eqref{action} and eventual renormalization of the field $\varphi$).

\section{One-loop renormalization of self-energy diagrams}\label{sec:1l2p}
In this section we analyze the one-loop two-point functions.
Quantum corrections generate a full propagator
\begin{align}\label{full-propagator}
    \langle\varphi_p\varphi_{p'}\rangle=\frac{\delta_{p+p'}}{p^2+m^2+\Sigma(p)}\,,
\end{align}
where $\Sigma(p)$ ---the self-energy of the scalar particle--- is given by the contributions of one-particle irreducible (1PI) two-point functions,
\begin{align}
    {\rm 1PI\ diagrams}=-\frac{\delta_{p+p'}}{(p^2+m^2)^2}\ \Sigma(p)\,.
\end{align}
We will analyze the perturbative structure of the self-energy,
\begin{align}
  \Sigma(p)=\hbar\,\Sigma_1(p)+\hbar^2\,\Sigma_2(p)+\ldots\,,
\end{align}
with particular emphasis in $d=2$ and $d=4$, in order to determine the kind of counterterms required by renormalization.

\begin{figure}[t]
\centering
\begin{minipage}{.4\textwidth}
\centering
\includegraphics[height=15mm]{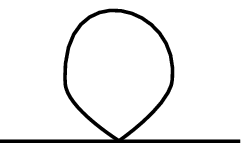}
\caption{\small One-loop planar contribution to the self-energy}
\label{2p1lp}
\end{minipage}
\hspace{0.05\textwidth}
\begin{minipage}{.4\textwidth}
\centering
\includegraphics[height=15mm]{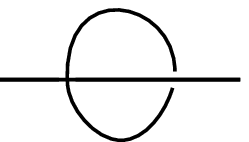}
\caption{\small One-loop nonplanar contribution to the self-energy}
\label{2p1lnp}
\end{minipage}
\end{figure}

One-loop contributions $\Sigma_1(p)$ to the self-energy $\Sigma(p)$ arise from all (connected) contractions between two external fields and the fields in the vertex \eqref{lambda}. In the commutative case all such contractions would give the same contribution because the vertex is invariant under any permutation of the internal momenta $k_1,\ldots,k_4$. However, this invariance is lost in the presence of the twisting factor  $\exp{i\pi(k_1\theta k_2+k_3\theta k_4)}$, which is only invariant under cyclic permutations of the internal momenta so there are three sets of four equivalent contractions. Since there are only two external fields, two of these sets of contractions give the same contribution due to momentum conservation. There are thus eight contractions which give the same contribution and a different type of contribution from the other four contractions. In terms of Feynman diagrams, the former are related to planar diagrams (Fig. \ref{2p1lp}) whereas the latter correspond to nonplanar ones (Fig. \ref{2p1lnp}). It is well known in noncommutative theories that the distinction between planar and nonplanar contributions plays a crucial role in the description of quantum corrections to any $n$-point function; the NC torus is not an exception to this fact.

As a consequence, $\Sigma_1(p)$ can be written as
\begin{align}\label{1ll}
    \Sigma_1(p)=8\lambda\,S_1(0)+4\lambda\,S_1(p)\,,
\end{align}
where $S_1$ represents the sum
\begin{align}
      S_1(p)=\sum_{k\in\mathbb{Z}^d}\frac{e^{2\pi i\,k\theta p}}{k^2+m^2}\,.\label{s1}
\end{align}
One can easily see that $S_1(p)$ is divergent for certain values of $p$ ---determined by the numerical character of $\theta$--- so we need an appropriate definition of this series that provides a regularization of its divergences.
In this article we regularize the divergencies of Feynman diagrams by introducing an arbitrary complex power $\epsilon$ of the free propagators (with ${\rm Re}(\epsilon)$ large enough),
\begin{align}
  \frac{1}{k^2+m^2}\rightarrow \frac{1}{\left(k^2+m^2\right)^{1+\epsilon}}\,,
\end{align}
and then performing the analytic extension to $\epsilon=0$; eventual divergencies then emerge as poles of this analytic extension. At some point, this technique can be related to dimensional regularization.

Let us study, in general, the sum
\begin{align}
      S_n(p,\epsilon)=\sum_{k\in\mathbb{Z}^d}\frac{e^{2\pi i\,k\theta p}}
      {\left\{(k^2+m^2)^n\right\}^{1+\epsilon}}\,,\label{sn}
\end{align}
whose analytic extension to $\epsilon=0$ for $n=1$ defines the expression $S_1(p)$ given in \eqref{s1}. If we introduce the Schwinger proper time representation we obtain
\begin{align}
      S_n(p,\epsilon)&=\frac{1}{\Gamma(n+n\epsilon)}
              \int_0^\infty dt\,t^{n+n\epsilon-1}\,e^{-tm^2}
              \sum_{k\in\mathbb{Z}^d}e^{-tk^2}e^{2\pi i\,k\theta p}\nonumber\\
                     &=\frac{\pi^{\frac{d}{2}}}{\Gamma(n+n\epsilon)}
              \int_0^\infty dt\,t^{n(1+\epsilon)-1-\frac{d}{2}}\,e^{-tm^2}
              \sum_{k\in\mathbb{Z}^d}e^{-\frac{\pi^2}{t}\,|k+\theta p|^2}
              \,.\label{snp}
\end{align}
In the last line of this expression we have used Poisson resummation,
\begin{align}
  \sum_{k\in \mathbb{Z}^d}f(k)
  =(2\pi)^d\sum_{k\in \mathbb{Z}^d}\tilde{f}(2\pi k)\,,
\end{align}
for $f(k)=\exp{(-tk^2+2\pi i k\theta p)}$ and $\tilde f$ its Fourier transform. It is convenient to consider separately the case in which $p\in\mathcal{Z}_\theta$, as defined in \eqref{theset}; recall that for rational $\theta$ this set is infinite, whereas for irrational $\theta$ the set $\mathcal{Z}_\theta$ is trivial. For $p\notin\mathcal{Z}_\theta$ each term in the sum of expression \eqref{snp} decreases exponentially for $t\rightarrow 0$ so the integration can be performed in the vicinity of $\epsilon=0$ and the result reads
\begin{align}\label{sn-inv}
  S_n(p,\epsilon)=\frac{2\pi^n}{(n-1)!}\,m^{\frac d2-n}\,\sum_{k\in\mathbb{Z}^d}
  \frac{K_{\frac{d}{2}-n}(2\pi m|k+\theta p|)}{|k+\theta p|^{\frac d2-n}}+O(\epsilon)\,,
\end{align}
where $K$ represents the modified Bessel function. The sum $S_1(p)$, originally defined in \eqref{s1}, is then given ---for $n=1$ and any $p\notin\mathcal{Z}_\theta$--- by the convergent series in the r.h.s.\ of \eqref{sn-inv}.

On the contrary, if the external momentum $p$ belongs to the set $\mathcal{Z}_\theta$, then the term in the series \eqref{snp} with $k=-\theta p$ does not present the exponential decrease for small $t$ so the integration must be performed for ${\rm Re}(\epsilon)>-1+d/2n$. If we separate this term we get, after integration in $t$,
\begin{align}
      S_n(p,\epsilon)=&\frac{\Gamma(n+n\epsilon-\frac{d}{2})}{\Gamma(n+n\epsilon)}\ \pi^{\frac d2}\,m^{d-2n-2n\epsilon}
              +\mbox{}\nonumber\\
              &\mbox{}+\frac{2\pi^{n+n\epsilon}m^{\frac d2-n-n\epsilon}}{\Gamma(n+n\epsilon)}
              \sum_{k\neq -\theta p}\frac{K_{n+n\epsilon-\frac{d}{2}}(2\pi m |k+\theta p|)}{|k+\theta p|^{\frac{d}{2}-n-n\epsilon}}
              \,.\label{sn0}
\end{align}
This expression shows that, for $p\in\mathcal{Z}_\theta$, the analytic extension of $S_n(p,\epsilon)$ has a simple pole at $\epsilon=0$ if $n\leq d/2$. In particular, for $n=1$ we obtain
\begin{align}
      S_1(p,\epsilon)=&-\frac12\,(-1)^{\frac d2}\,V_d\ m^{d-2}\ \frac1{\epsilon}
      +\frac12\,(-1)^{\frac d2}\,V_d\ m^{d-2}\left\{\log{m^2}-\psi\left(\tfrac d2\right)
      -\gamma\right\}+\mbox{}\nonumber\\[2mm]
      &\mbox{}+2\pi\, m^{\frac d2-1}
      \sum_{k\neq -\theta p}\frac{K_{\frac{d}{2}-1}(2\pi m |k+\theta p|)}{|k+\theta p|^{\frac{d}{2}-1}}
      +O(\epsilon)\,,\label{s10}
\end{align}
where $V_d$ is the volume of the sphere $S^{d-1}$. Therefore, the original sum \eqref{s1} can be written, for $p\in\mathcal{Z}_\theta$, as
\begin{align}
  S_1(p)=-\frac12\,(-1)^{\frac d2}\,V_d\ m^{d-2}\ \frac1{\epsilon}+({\rm finite\ terms})\,.
\end{align}
In conclusion, $S_1(p)$ is conditionally convergent for $p\notin\mathcal{Z}_\theta$ but diverges as $\sim m^{d-2}/\epsilon$ otherwise, in particular for $p=0$.

The divergent contribution of $S_1(0)$ to $\Sigma_1(p)$ (see \eqref{1ll}) can be removed by an ordinary mass redefinition (see \eqref{full-propagator}),
\begin{align}\label{mass-div}
  m^2\rightarrow m^2\ \left(1+8\lambda\ \frac{(-\pi)^{\frac d2}}{\Gamma\left(\frac d2\right)}
  \ \frac{m^{d-4}}{\epsilon}\right)\,.
\end{align}
However, due to the term $S_1(p)$ in \eqref{1ll}, $\Sigma_1(p)$ might still be divergent if the external momentum $p$ belongs to $\mathcal{Z}_\theta$ so we need to introduce new mass terms in the action
\begin{align}
  \tfrac12\,\mu^2\,\sum_{p\in\mathcal{Z}_\theta}\left|\varphi_p\right|^2\,,\label{mu}
\end{align}
with
\begin{align}\label{mu-div}
  \mu^2=4\lambda\,\frac{(-\pi)^{\frac d2}}{\Gamma\left(\frac d2\right)}
  \ \frac{m^{d-2}}{\epsilon}\,,
\end{align}
for those field components $\varphi_p$ such that $p\in\mathcal{Z}_\theta$. This is one of the counterterms present in expression \eqref{ctaction}.

In consequence, after appropriate $O(\lambda)$ mass renormalizations, $\Sigma_1(p)$ is finite for any value of $p$. Note that, upon quantum corrections, the mass of the field takes a different value for field components with momentum in $\mathcal{Z}_\theta$. In particular, for irrational $\theta$ the new term \eqref{mu} in the action can be written as
\begin{align}
  \tfrac12\,\mu^2\,\tau(\varphi)\,\tau(\varphi)\,,\label{mudiop}
\end{align}
so only the zero-momentum component $\varphi_0$ of the field gets a different mass.

Although the one-loop correction to the self-energy for any value of the external momentum is rendered finite by the mass renormalizations, the sum of all these contributions ---implicit in the effective action--- can be seen to be convergent, for irrational $\theta$, only under the Diophantine condition \cite{Gayral:2006vd}.

Having computed these corrections, we want to analyze the dependence of two-point functions with the numerical character of $\theta$. Let us then consider two noncommutativity matrices, $\theta_1$ and $\theta_2$, one being rational while the other - irrational. Even though the difference $|| \theta_1 -\theta_2||$ may be arbitrarily small, the counterterms vary drastically from $\theta_1$ to $\theta_2$. This large variation is pretty harmless if it can be removed from the amplitudes by a finite renormalization of couplings. Let us see if this is the case at the example of the one-loop two-point function in $d=2$. Let $\theta_1$ be rational, and $\theta_2$ be a Diophantine noncommutativity matrix very close to $\theta_1$. The planar diagram does no depend on $\theta$, so that we shall consider nonplanar contributions only. Let us take $p\in \mathcal{Z}_{\theta_1}\backslash \{ 0 \}$. If apart from the pole term in \eqref{mu-div} one allows for a finite renormalization of $\mu^2$, the finite part of $S_1(p,\epsilon)$ may be shifted to an arbitrary $p$-independent value. Therefore, the renormalized two-point function reads
\begin{equation}
S_1^R(p)_{\theta_1} = s_1 + 2\pi \sum_{k\ne -\theta_1 p} K_0(2\pi m|k+\theta_1p|) \,,\label{S11}
\end{equation}
where $s_1$ has to be fixed by a suitable normalization condition. Since $\theta_2$ is Diophantine, $S_1(p,\epsilon)_{\theta_2}$ is not divergent, and its renormalized value is just the $\epsilon\to 0$ limit of \eqref{sn-inv},
\begin{equation}
S_1^R(p)_{\theta_2} =2\pi K_0(2\pi m|(\theta_1-\theta_2)p|)+2\pi \sum_{k\ne -\theta_1 p} K_0(2\pi m|k+\theta_2p|)\,,\label{S12}
\end{equation}
where we separated one of the terms in the infinite sum in \eqref{sn-inv}. Consider $S_1^R(p)_{\theta_1}-S_1^R(p)_{\theta_2}$ in the limit $||\theta_1-\theta_2||\to 0$. The contributions of $k\ne -\theta_1 p$ cancel in this limit, as one can easily see. $s_1$ may depend on $\theta_1$, but definitely not on $\theta_2$. Therefore,
\begin{equation}
S_1^R(p)_{\theta_1}-S_1^R(p)_{\theta_2} = -2\pi K_0(2\pi m|(\theta_1-\theta_2)p|) + O(1)=
2\pi \ln |(\theta_1-\theta_2)p| +O(1). \label{S1112}
\end{equation}
Hence, the variation of two-point function grows indefinitely as $\theta_2$ approaches $\theta_1$. Some implications of this result will be discussed below in Sect.\ \ref{sec:con}.
Note that since $\theta_1\ne \theta_2$ both two-point functions, $S_1^R(p)_{\theta_1}$ and $S_1^R(p)_{\theta_2}$, are always finite.

One may find some similarities between this situation and the one in the matrix model approach to noncommutivity, where the effective action behaves quite irregularly for some relations between parameters of the theory \cite{Blaschke:2010rr}.
\section{One-loop renormalization of four-point functions}\label{four-point}

\begin{figure}[t]
\centering
\begin{minipage}{.4\textwidth}
\centering
\includegraphics[height=12mm]{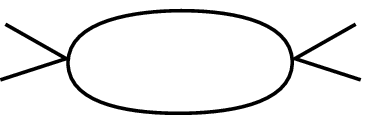}
\caption{\small Planar contribution to the four-point function}
\label{4pfish}
\end{minipage}
\hspace{0.05\textwidth}
\begin{minipage}{.4\textwidth}
\centering
\includegraphics[height=12mm]{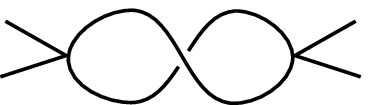}
\caption{\small Nonplanar contribution to the four-point function}
\label{4pfishnp}
\end{minipage}
\pagebreak
\vspace{.05\textwidth}
\begin{minipage}{.4\textwidth}
\centering
\includegraphics[height=12mm]{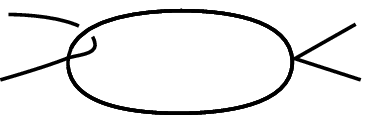}
\caption{\small Nonplanar contribution to the four-point function}
\label{4p1t}
\end{minipage}
\hspace{0.05\textwidth}
\begin{minipage}{.4\textwidth}
\centering
\includegraphics[height=12mm]{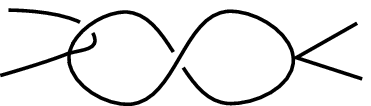}
\caption{\small Nonplanar contribution to the four-point function}
\label{4p1tnp}
\end{minipage}
\pagebreak
\vspace{.05\textwidth}
\begin{minipage}{.4\textwidth}
\centering
\includegraphics[height=12mm]{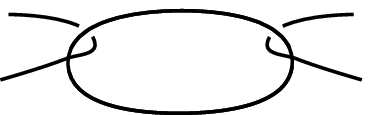}
\caption{\small Nonplanar contribution to the four-point function}
\label{4p2t}
\end{minipage}
\hspace{0.05\textwidth}
\begin{minipage}{.4\textwidth}
\centering
\includegraphics[height=12mm]{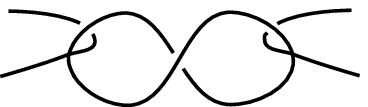}
\caption{\small Nonplanar contribution to the four-point function}
\label{4p2tnp}
\end{minipage}
\end{figure}

In order to complete the analysis of one-loop divergencies we consider the four-point function with external momenta $p_1,p_2,p_3,p_4$. The contributions of the different Feynman diagrams to the $s$-channel ($p_1,p_2$ entering the same vertex) are given by
\begin{align}
  &64\,\lambda^2\,e^{\pi i(p_1\theta p_2+p_3\theta p_4)}\,L(p_1+p_2,0)\,,
    &({\rm Figure\ }\ref{4pfish})\label{1}\\
  &64\,\lambda^2\,e^{\pi i(p_1\theta p_2+p_3\theta p_4)}\,L(p_1+p_2,p_1+p_2)\,,
    &({\rm Figure\ }\ref{4pfishnp})\label{2}\\
  &32\,\lambda^2\,e^{\pi i(p_1\theta p_2+p_3\theta p_4)}\,\left[L(p_1+p_2,p_2)+L(p_1+p_2,p_4)\right]\,,
    &({\rm Figure\ }\ref{4p1t})\label{3}\\
  &32\,\lambda^2\,e^{\pi i(p_1\theta p_2+p_3\theta p_4)}\,\left[L(p_1+p_2,p_1)+L(p_1+p_2,p_3)\right]\,,
    &({\rm Figure\ }\ref{4p1tnp})\label{4}\\
  &16\,\lambda^2\,e^{\pi i(p_1\theta p_2+p_3\theta p_4)}\,L(p_1+p_2,p_1+p_4)\,,
    &({\rm Figure\ }\ref{4p2t})\label{5}\\
  &16\,\lambda^2\,e^{\pi i(p_1\theta p_2-p_3\theta p_4)}\,L(p_1+p_2,p_1+p_3)\,.
    &({\rm Figure\ }\ref{4p2tnp})\label{6}
\end{align}
In these expressions $L(p,q)$ is defined as the analytic extension to $\epsilon=0$ of
\begin{align}
      L(p,q,\epsilon)=\sum_{k\in\mathbb{Z}^d}
      \frac{e^{2\pi i\,k\theta q}}{\left\{[(k+p)^2+m^2](k^2+m^2)\right\}^{1+\epsilon}}\,.\label{lpq}
\end{align}
The two terms in \eqref{3} and \eqref{4} correspond to the cases where the incoming momentum $p_1+p_2$ enters the diagrams from the left or from the right.

In order to study the analytic extension of the sum \eqref{lpq} we introduce Feynman parameters $u,v$ to collect both propagators into a single denominator, we use the Schwinger proper time representation and then the Poisson resummation formula; the result reads
\begin{align}\label{ele}
        L(p,q,\epsilon)&=\sum_{k\in\mathbb{Z}^d}e^{2\pi i\,k\theta q}
        \ \frac{\Gamma(2+2\epsilon)}{\Gamma^2(1+\epsilon)}
        \int_0^1\int_0^1 du\,dv
        \ \frac{\delta(u+v-1)\,(uv)^\epsilon}{\left\{k^2+2ukp+up^2+m^2\right\}^{2+2\epsilon}}\nonumber\\
        &=\frac{\pi^\frac d2}{\Gamma^2(1+\epsilon)}
        \int_0^1 du\,[u(1-u)]^\epsilon
        \int_0^\infty dt\,t^{1-\frac d2+2\epsilon}\,e^{-t[m^2+u(1-u)p^2]}\times\mbox{}\nonumber\\
        &\mbox{}\times\sum_{k\in\mathbb{Z}^d}
        e^{-\frac{\pi^2}{t}|k+\theta q|^2-2\pi i u p(k+\theta q)}\,.
\end{align}
As before, if $q\notin\mathcal{Z}_\theta$ then each term in the series is exponentially decreasing for small $t$ so it can be integrated in some neighborhood of $\epsilon=0$; $L(p,q)$ is thus finite for $q\notin\mathcal{Z}_\theta$. On the other hand, for $q\in\mathcal{Z}_\theta$, integration in $t$ gives
\begin{align}\label{elep}
        L(p,q,\epsilon)=&\frac{(-\pi)^\frac d2}{\Gamma\left(\frac d2-1\right)}
        \ \int_0^1 du\, \left[m^2+u(1-u)p^2\right]^{\frac d2-2}\times\mbox{}\nonumber\\
        &\mbox{}\times\left\{\ \frac1{2\epsilon}
        -\log{\left[m^2+u(1-u)p^2\right]}+\log{\sqrt{u(1-u)}}+\gamma+\psi(\tfrac{d}{2}-1)\right\}
        +\mbox{}\nonumber\\
        &\mbox{}+2\pi^2\int_0^1 du\,\left[m^2+u(1-u)p^2\right]^{\frac d4-1}\times\mbox{}\nonumber\\
        &\mbox{}\times\sum_{k\neq -\theta q}\frac{e^{-2\pi i u p(k+\theta q)}}{|k+\theta q|^{\frac d2-2}}
        \ K_{d/2-2}\left(2\pi|k+\theta q|\sqrt{m^2+u(1-u)p^2}\right)
        +O(\epsilon)\,.
\end{align}
The sums $L(p,q)$ that determine the contributions of the diagrams displayed in figures \ref{4pfish}--\ref{4p2tnp} can then be written, for $q\in\mathcal{Z}_\theta$, as
\begin{align}\label{divL}
        L(p,q)&=\frac{(-\pi)^\frac d2}{\Gamma\left(\frac d2-1\right)}
        \ \int_0^1 du\, \left(m^2+u(1-u)p^2\right)^{\frac d2-2}\ \frac1{2\epsilon}
        +({\rm finite\ terms})\,.
\end{align}
This expression is finite for $d=2$ (with a branch cut at $p^2=4m^2$) and diverges as $\pi^2/2\epsilon$ (independently of $p$) for $d=4$. In higher dimensions the residue depends on $p$.

Let us therefore analyze the counterterms that are needed in four dimensions in order to remove the resulting divergencies of the four-point functions. The contribution \eqref{1} ---corresponding to the planar diagram in Fig. \ref{4pfish}--- contains an UV divergence, which can be removed by a renormalization of the self-coupling constant,
\begin{align}\label{ctl}
  \lambda &\rightarrow \lambda\left(1+4\pi^2\lambda\ \frac{1}{\epsilon}\right)\,.
\end{align}
Besides, contributions \eqref{3} and \eqref{4} ---corresponding to the diagrams in Figs. \ref{4p1t} and \ref{4p1tnp}--- together with the t- and u-channels are also divergent if any of the incoming momenta belongs to the set $\mathcal{Z}_\theta$. This type of divergence can be removed by introducing the following self-interaction, corresponding to the second term in \eqref{ctaction}:
\begin{align}
    \lambda_1\,
    \sum_{k\in\mathcal{Z}_\theta}\varphi_k
    \sum_{k_1,k_2,k_3 \in\mathbb{Z}^d}\delta_{k+k_1+k_2+k_3}\,\varphi_{k_1}\varphi_{k_2}\varphi_{k_3}
    \,e^{i\pi\,k_1\theta k_2+i\pi\,k_3\theta k}\,,\label{lambda1}
\end{align}
with
\begin{align}\label{ctl1}
  \lambda_1=8\pi^2\lambda^2\ \frac{1}{\epsilon}\,.
\end{align}
For irrational $\theta$, this new interaction reads
\begin{align}
    \lambda_1\,\tau(\varphi)\,\tau(\varphi^3)\,.\label{lambda1diop}
\end{align}
Lastly, contributions \eqref{2}, \eqref{5} and \eqref{6} ---corresponding to the diagrams in Figs. \ref{4pfishnp}, \ref{4p2t} and \ref{4p2tnp}--- present a divergence whenever the sum of two incoming momenta belongs to $\mathcal{Z}_\theta$ whose cancellation requires the following self-interaction, corresponding to the third term in \eqref{ctaction}:
\begin{align}
    \lambda_2\,
    \sum_{k\in\mathcal{Z}_\theta}\sum_{k_1,\ldots,k_4}
    \delta_{k_1+k_2-k}\,\delta_{k_1+\ldots+k_4}\,
    \varphi_{k_1}\varphi_{k_2}\varphi_{k_3}\varphi_{k_4}
    \,e^{i\pi\,k_1\theta k_2+i\pi\,k_3\theta k}\,,\label{lambda2}
\end{align}
with
\begin{align}\label{ctl2}
  \lambda_2=6\pi^2\lambda^2\ \frac{1}{\epsilon}\,.
\end{align}
For irrational $\theta$, the counterterm \eqref{lambda2} reads
\begin{align}
    \lambda_2\,\tau(\varphi^2)\,\tau(\varphi^2)\,.\label{lambda2diop}
\end{align}
After the introduction of these counterterms four-point functions in $\mathbb{T}^4_\theta$ are rendered finite for any value of the external momenta. Note that all $\beta$-functions associated with the coupling constants $\lambda,\lambda_1,\lambda_2$ are positive.

\section{Higher loops at two dimensions}\label{sec-2d}
Before analyzing higher order of perturbation series on $\mathbb{T}^2_\theta$ let us briefly recall the UV/IR mixing problem on noncommutative plane $\mathbb{R}^d_\theta$. The nonplanar diagrams on $\mathbb{R}^d_\theta$ behave better in the ultraviolet than their commutative counterparts since $(p\theta)^{-1}$ (with $p$ being an external momentum) serves as an effective ultraviolet cutoff. However, the divergences are restored in the commutative limit, $\theta\to 0$, implying also a singularity at $p\to 0$. According to Ref.\ \cite{Minwalla:1999px}, these singularities cause troubles with the convergence of loop integrals at zero momenta if nonplanar diagrams are inserted into internal lines of other diagrams. Note that in two dimensions 1PI diagrams are at most logarithmically divergent, so that the IR singularity may also be at most $\ln |p|$. This singularity is rather mild. Thus one does not expect much troubles with the UV/IR mixing in $d=2$. For this reason, our consideration of the two-torus will also be rather sketchy.

Turning to $\mathbb{T}^2_\theta$, we first note that there are only two diagrams, Figs.~\ref{2p1lp} and \ref{2p1lnp}, that are superficially divergent. By using the expression \eqref{s10} and basic properties of $K_0$, one can easily show that after adding the counterterm from \eqref{mu} the nonplanar diagram Fig.~\ref{2p1lnp} with $p \in \mathcal{Z}_\theta$ becomes a bounded function of $p$. For $p\not\in\mathcal{Z}_\theta$, there may be a growing contribution to $S_1(p)$, which comes from the momentum $k_p$ in \eqref{sn0} that minimizes $|k+\theta p|$. It reads
\begin{equation}
2\pi K_0(2\pi m|k_p+\theta p|) \simeq -2\pi \ln |k_p+\theta p| \,.\label{kp}
\end{equation}
By the Diophantine condition \eqref{Dioph}, this contribution is restricted by $2\pi (1+\beta)\ln |p|$ at large $|p|$. Therefore, the renormalized 2-point function on $\mathbb{T}^2_\theta$ has a logarithmic singularity, but in contrast to $\mathbb{R}^2_\theta$ this singularity is UV rather than IR. The UV singularities on the quantum plane, discussed in \cite{Chaichian:1999wy}, were found more severe than on the commutative plane. However, the singularities on $T^2_\theta$ are very mild. Indeed, if one inserts the renormalized diagram of Fig.\ \ref{2p1lnp} into an internal line with the momentum $p$ of some other diagram one gets (at large $|p|$) a multiplier of $\ln |p|$ from the diagram itself and $(p^2+m^2)^{-1}$ from an extra propagator. The overall contribution behaves as $\ln |p| \cdot (p^2+m^2)^{-1}$ and does even improve the convergence of larger diagram.

We saw that in the $\varphi^4$ theory on $\mathbb{T}^2_\theta$ (i) all superficially divergent diagrams can be renormalized by the counterterms that we have proposed, and (ii) the insertion of renormalized superficially divergent diagrams does not lead to any problems with convergence. Hence, there is no UV/IR mixing in this model, and it is likely renormalizable.
\section{Two-loop self-energy at four dimensions}\label{sec:2l4D}

In this last section we describe the diagrams that contribute to $\Sigma_2(p)$ ---the second order correction to the self-energy--- in the four-dimensional torus. For corresponding analysis on $\mathbb{R}^4_\theta$ one may consult Ref.\ \cite{Micu:2000xj}. In this section we restrict ourselves to the case of a pure irrational (Diophantine) noncommutativity parameter. Therefore, $\mathcal{Z}_\theta =\{ 0\}$. We analyze the divergences of two-loop diagrams and point out the main difficulties one finds in computing the remaining double sums; in the course of this analysis we will see the importance of the Diophantine condition on the matrix $\theta$. Let us also remark that, since we are interested in the renormalizability of two-point functions, we will neglect divergent contributions which are either independent or quadratic in the external momentum $p$ for they can be removed by mass and field renormalizations of order $O(\lambda^2)$.

Before considering two-loop diagrams, we analyze the $O(\lambda^2)$ contributions of the counterterms already introduced in the previous sections, i.e.\ one-loop diagrams built with the leading quantum corrections to the parameters $m^2,\lambda$, as well as with the new parameters $\mu^2,\lambda_1,\lambda_2$. In the first place, the nonplanar tadpole in Fig. \ref{2p1lnp} gives an $O(\lambda^2)$ contribution,
\begin{align}\label{lambda2m}
  -32\,\lambda^2m^2\,S_2(p)\ \frac{\pi^2}{\epsilon}\,,
\end{align}
from the insertion of the mass correction \eqref{mass-div} into its  internal propagator, as well as another $O(\lambda^2)$ contribution,
\begin{align}\label{lambda2l}
  16\,\lambda^2\,S_1(p)\ \frac{\pi^2}{\epsilon}\,,
\end{align}
from the insertion of the $\lambda$ correction \eqref{ctl} into its vertex. In \eqref{lambda2m} $S_2(p)$ is defined as the analytic extension to $\epsilon=0$ of \eqref{sn} (for $n=2$ and $d=4$). Second, a planar tadpole of the type shown in Fig. \ref{2p1lp} at the vertex $\lambda_2$ gives
\begin{align}\label{lambda2l2}
  48\,\lambda^2\,\frac 1{p^2+m^2} \ \frac{\pi^2}{\epsilon}\,.
\end{align}

Equations \eqref{lambda2m}, \eqref{lambda2l} and \eqref{lambda2l2} represent nonlocal (and not trace-like) divergencies introduced in the self-energy by the renormalization of the parameters in the action; contributions of $\mu^2$ and $\lambda_1$ are not taken into account for they are either $p$-independent or quadratic in $p$.
\begin{figure}[t]
\centering
\begin{minipage}{.4\textwidth}
\centering
\includegraphics[height=20mm]{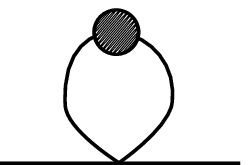}
\caption{\small Planar diagram with a $\Sigma_1(p)$ insertion}
\label{PlanSigma}
\end{minipage}
\hspace{0.05\textwidth}
\begin{minipage}{.4\textwidth}
\centering
\includegraphics[height=20mm]{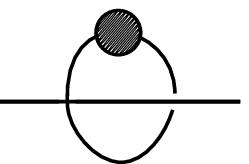}
\caption{\small Nonplanar diagram with a $\Sigma_1(p)$ insertion}
\label{NonPlanSigma}
\end{minipage}
\end{figure}

\begin{figure}[t]
\centering
\begin{minipage}{.42\textwidth}
\centering
\includegraphics[height=12mm]{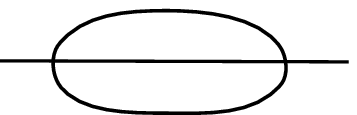}
\caption{\small Planar diagram contributing with $-16\lambda^2\,U(p,0)$ to $\Sigma_2(p)$}
\label{2l-sunset}
\end{minipage}
\hspace{0.05\textwidth}
\begin{minipage}{.4\textwidth}
\centering
\includegraphics[height=12mm]{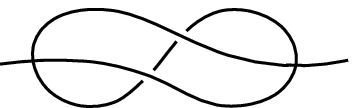}
\caption{\small Nonplanar diagram contributing with $-32\lambda^2\,U(p,p)$ to $\Sigma_2(p)$}
\label{2l-scalex}
\end{minipage}
\pagebreak
\vspace{.05\textwidth}
\begin{minipage}{.4\textwidth}
\centering
\includegraphics[height=12mm]{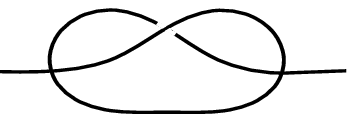}
\caption{\small Nonplanar diagram contributing with $-32\lambda^2\,V(p,0)$ to $\Sigma_2(p)$}
\label{2l-brazos}\end{minipage}
\hspace{0.05\textwidth}
\begin{minipage}{.4\textwidth}
\centering
\includegraphics[height=12mm]{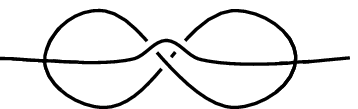}
\caption{\small Nonplanar diagram contributing with $-16\lambda^2\,V(p,p)$ to $\Sigma_2(p)$}
\label{2l-lentes}
\end{minipage}
\end{figure}

The two-loop diagrams can be built in two different ways: either by inserting the one-loop self-energy $\Sigma_1(p)$ into the internal propagator of a planar or a nonplanar tadpole (see Figs. \ref{PlanSigma},\ref{NonPlanSigma}, respectively), so that both external legs enter the same vertex, or by attaching each external leg to a different vertex so that the two loops share a common internal momentum, as in Figs. \ref{2l-sunset}, \ref{2l-scalex}, \ref{2l-brazos} and \ref{2l-lentes}.

For the first type of diagrams, the insertion into a planar tadpole, Fig.~\ref{PlanSigma}, gives a contribution which, though divergent, does not depend on the external momentum. As we have already explained above, such contributions are harmless and will be discarded.
The contributions corresponding into the insertion into a nonplanar tadpole, Fig.~\ref{NonPlanSigma}, are given, up to $O(\lambda^2)$, by
\begin{align}
  -4\lambda\sum_{k\in\mathbb{Z}^4}\frac{e^{2\pi i\,k\theta p}}{(k^2+m^2)^2}\,\Sigma_1(k)\,.\label{1tlambdalambda}
\end{align}
Replacing \eqref{1ll} into this expression we obtain
\begin{align}
  -32\lambda^2\,S_1(0)S_2(p)-16\lambda^2\,T(p)\,,\label{1t}
\end{align}
where $T(p)$ is defined as the analytic extension to $\epsilon=0$ of the series
\begin{align}
        T(p,\epsilon)=\sum_{k,l\in\mathbb{Z}^4}
        \frac{e^{2\pi i\,k\theta l}}{\{[(k+p)^2+m^2](l^2+m^2)^2\}^{1+\epsilon}}\,.\label{tp}
\end{align}
Note that the first term in \eqref{1t}, though $p$-dependent, is exactly cancelled by \eqref{lambda2m}.

Lastly, the contributions of diagrams which contain overlapping divergencies read (see Figs. \ref{2l-sunset}, \ref{2l-scalex}, \ref{2l-brazos}, \ref{2l-lentes}),
\begin{align}\label{2tlambdalambda-div}
  -16\lambda^2\,U(p,0)-32\lambda^2\,U(p,p)-32\lambda^2\,V(p,0)-16\lambda^2\,V(p,p)\,,
\end{align}
where the sums $U,V$ are defined as the analytic extensions to $\epsilon=0$ of the series
\begin{align}
    U(p,q,\epsilon)&=\sum_{k,l\in\mathbb{Z}^4}
            \frac{e^{2\pi i\,l\theta q}}
            {\{[(k+p)^2+m^2](l^2+m^2)[(l+k)^2+m^2)]\}^{1+\epsilon}}\,,\label{upq}\\
    V(p,q,\epsilon)&=\sum_{k,l\in\mathbb{Z}^4}
            \frac{e^{2\pi i\,k\theta q}\,e^{2\pi i\,k\theta l}}
            {\{[(k+p)^2+m^2](l^2+m^2)[(l+k)^2+m^2)]\}^{1+\epsilon}}\,.\label{vpq}
\end{align}
Since $U(p,0)$ corresponds to the diagram of the ordinary commutative case, its divergence is a quadratic polynomial in $p$. According to App. \ref{ap_upp}, the sum $U(p,p,\epsilon)$ behaves as
\begin{align}
  -32\lambda^2\,U(p,p,\epsilon)=-16\,\lambda^2\,S_1(p)\ \frac{\pi^2}{\epsilon}
  +{\rm quad.\, pol.}+O(\epsilon)\,,
\end{align}
where we have represented by ``quad.\,pol.''\ terms which, though eventually divergent, are quadratic expressions\footnote{More precisely, ``quad.\ pol.''\ should have the form $a+bp^2$, i.e. the coefficient in front of $p_\mu p_\nu$ has to be proportional to the unit matrix.} in $p$. Therefore, the nonlocal non-trace-like divergence introduced by $U(p,p)$ completely cancels \eqref{lambda2l}.

In consequence, two-loop renormalization of the self-energy demands that the remaining nonlocal divergencies (given by \eqref{lambda2l2}, the second term in \eqref{1t} and the last two terms in \eqref{2tlambdalambda-div}) cancel among each other. In other words, the remaining potentially divergent terms read
\begin{align}\label{remainingD}
  48\,\lambda^2\,\frac 1{p^2+m^2}\ \frac{\pi^2}{\epsilon}
  -16\lambda^2\,T(p,\epsilon)
  -32\lambda^2\,V(p,0,\epsilon)
  -16\lambda^2\,V(p,p,\epsilon)\,.
\end{align}
The divergences of this expression at $\epsilon=0$ must repeat the structure of quadratic (in $\varphi$) counterterms. I.e., they have to be of the form of a quadratic polynomial in $p$ plus a term proportional to $\delta_p$.

The double sums in \eqref{remainingD} can all be treated in a unified way. The divergences of these sums in $\mathbb{Z}^8$ at $\epsilon=0$ arise from the fact that the denominator increases only with a sixth power at infinity. Nevertheless, the twisting factor $e^{2\pi i\,k\theta l}$ contributes to regularize the series. This certainly happens in the continuum case where the corresponding integration in $\mathbb{R}^8$ is finite. However, in the discrete case there exist four-dimensional subspaces ---with null measure in $\mathbb{R}^8$--- for which the twisting factor vanishes.

Therefore, the divergencies of $T(p,\epsilon)$ can be attributed to the subseries in the subspaces for which $k\theta l=0$ and, simultaneously, the denominator increases at infinity with a power which is less or equal than four. Such subseries correspond to $l=0$ and
$k=0$. Thus,
\begin{equation}
T(p,\epsilon)=\sum_{l\in \mathbb{Z}^4} \frac 1{\{(l^2+m^2)^2(p^2+m^2)\}^{1+\epsilon}}+
\sum_{k\in \mathbb{Z}^4} \frac 1{\{((k+p)^2+m^2)^2m^4\}^{1+\epsilon}}+O(1)\,.\label{Tpe}
\end{equation}
The second sum in \eqref{Tpe} does not depend on $p$, while the first one is proportional to $S_2(0,\epsilon)$; see \eqref{sn}. We conclude that
\begin{align}
  T(p,\epsilon)&=\frac{1}{p^2+m^2}\ \frac{\pi^2}{2\epsilon}+{\rm quad.\, pol.}+O(1)\,.\label{tcon}
\end{align}
The same formula is reobtained in App. \ref{ap_tp} by using mathematically rigorous methods.

To analyze $V(p,q,\epsilon)$, let us first change the summation variables, so that
\begin{align}\label{vpp_ap}
    V(p,q,\epsilon)&=\sum_{k,l\in\mathbb{Z}^4}
            \frac{e^{2\pi i\,k\theta l}}
            {\{[(k+p)^2+m^2][(l-q)^2+m^2][(l+k-q)^2+m^2)]\}^{1+\epsilon}}\,.
\end{align}
Using the same argument as above, potentially divergent terms come from the subsets $k=0$, $l=0$ and $k+l=0$, which contribute as
\begin{align}\label{ttd}
    &\frac{1}{(p^2+m^2)^{1+\epsilon}}\ S_2(0,\epsilon)
    +\frac{1}{(q^2+m^2)^{1+\epsilon}}\ L(p+q,0,\epsilon)
    +\frac{1}{(q^2+m^2)^{1+\epsilon}}\ L(p-q,0,\epsilon)\,.
\end{align}
The remaining terms in the series \eqref{vpp_ap} are expected to decrease for large $k,l$ as long as $|k-\theta l|$ does not decrease too fast, which is guaranteed by the Diophantine condition on $\theta$. This implies in particular
\begin{align}
  V(p,0,\epsilon)&=\frac{1}{p^2+m^2}\ \frac{\pi^2}{2\epsilon}+{\rm quad.\, pol.}+O(1)\,,\label{vpcon}\\
  V(p,p,\epsilon)&=3\ \frac{1}{p^2+m^2}\ \frac{\pi^2}{2\epsilon}+{\rm quad.\, pol.}+O(1)\,.\label{vppcon}
\end{align}

Unfortunately, we cannot reconfirm \eqref{vpcon} and \eqref{vppcon} by more rigorous methods.

From Eqs. \eqref{tcon}, \eqref{vpcon} and \eqref{vppcon} one concludes that \eqref{remainingD} does not contain any divergences except for the ones that can be removed by a renormalization of couplings in the action
\begin{align}
    S[\varphi]=&\tfrac 12\,\tau(\partial\varphi\,\partial\varphi)
    +\tfrac12\,m^2\,\tau(\varphi^2)
    +\lambda\,\tau(\varphi^4)+\mbox{}\nonumber\\[2mm]
    &\mbox{}+\tfrac12\,\mu^2\,\tau(\varphi)\,\tau(\varphi)
    +\lambda_1\,\tau(\varphi)\,\tau(\varphi^3)
    +\lambda_2\,\tau(\varphi^2)\,\tau(\varphi^2)\,.
\end{align}
Let us recall that in this section we assumed that $\theta$ is irrational.

Although our analysis is far from being exhaustive, we believe it strongly suggest that the $\varphi^4$ theory on $\mathbb{T}_\theta^4$ is renormalizable.

\section{Conclusions}\label{sec:con}

In this paper, we analyzed the renormalization of a scalar field theory on $\mathbb{T}_\theta^d$ with a quartic self-interaction after the introduction of a new type of nonlocal (but trace-like) interactions suggested by the previous heat kernel calculations \cite{Gayral:2006vd}. At one loop our analysis is complete. We also argued that in two dimensions no problems appear at higher orders as well. In four dimensions, we checked the renormalization of self-energy at two-loop order relying on our understanding of the behavior of double sums, which we were able to reconfirm by rigorous methods for all diagrams but one.  Our findings strongly suggest that the $\varphi^4$ theory on $\mathbb{T}^2_\theta$ and $\mathbb{T}^4_\theta$ is renormalizable.  The renormalization always strongly depends on the Diophantine character of the noncommutativity matrix $\theta$. We cannot exclude completely that some more elaborate multiple-trace counterterms will be needed, though their algebraic nature is less clear than that of the ones listed \eqref{ctaction}. To check this, one has to calculate the two-loop four-point functions.

On the technical side, it is important to develop the theory of regularized multiple sums with twisting factors. To the best of our knowledge, such sums have not been considered in the mathematics literature so far (see, e.g. \cite{Paycha}).

While calculating the renormalized two-point functions we encountered a potentially troubling phenomenon: these functions depend too strongly on the noncommutativity matrix. In other words, an arbitrarily small error in $\theta$ may cause an arbitrarily large variation in the two-point functions. Or, the value of two-point functions cannot be predicted unless we know $\theta$ with an infinite precision. This does not necessarily mean, however, that the theory is meaningless. We can suggest the following explanations and ways to overcome the difficulty.

\begin{enumerate}
\item Since the plane waves $U_p$ do not commute even classically, see \eqref{Upq}, they probably do not form a good basis. Therefore, the correlation functions of plane waves may be of little physical relevance by themselves. The problem is then to find a physically motivated basis of states that will ensure a kind of ``smooth'' dependence of the correlation functions on $\theta$.
\item One can try to achieve a meaningful answer by smearing the correlation functions over a small vicinity of a given $\theta$. The key issue is to find an appropriate measure.
\item Finally, perhaps one can extend the model to fix $\theta$ sharply to certain value, e.g. by some topological considerations.
\end{enumerate}
Although we cannot offer much details on any of the items above,  we believe that these directions deserve further study.

\section*{Acknowledgements}
DVV was supported in part by FAPESP, Project 2012/00333-7,  CNPq, Project 306208/2013-0 and by the Tomsk State University Competitiveness Improvement Program. DD acknowledges support from CONICET and UNLP (Proj. 11/X615), Argentina. PP acknowledges support from CONICET (PIP 1787/681), ANPCyT (PICT-2011-0605) and UNLP (Proj. 11/X615), Argentina.

\appendix

\section{The double sum $U(p,p)$}\label{ap_upp}

Let us consider the analytic extension to $\epsilon=0$ of the sum
\begin{align}\label{upp}
    U(p,p,\epsilon)&=\sum_{k,l\in\mathbb{Z}^4}\frac{e^{2\pi i\,l\theta p}}
    {\{[(k+p)^2+m^2](l^2+m^2)[(l+k)^2+m^2)]\}^{1+\epsilon}}\nonumber\\
    &=\sum_{l\in\mathbb{Z}^4}\frac{e^{2\pi i\,l\theta p}}{[(l+p)^2+m^2]^{1+\epsilon}}
    \,L(l,0,\epsilon)\,.
\end{align}
Using \eqref{ele} at $d=4$, this expression reads
\begin{align}
    U(p,p,\epsilon)=&\frac{\pi^2}{\Gamma^2(1+\epsilon)}
    \sum_{l\in\mathbb{Z}^4}\frac{e^{2\pi i\,l\theta p}}{[(l+p)^2+m^2]^{1+\epsilon}}
    \int_0^1 du\,[u(1-u)]^\epsilon\times\mbox{}\nonumber\\
    &\mbox{}\times\int_0^\infty dt\,t^{-1+2\epsilon}\,e^{-t(m^2+u(1-u)l^2)}
    \,\sum_{k\in\mathbb{Z}^4}e^{-\frac{\pi^2}{t}k^2-2\pi i u lk}
    \,.
\end{align}
If we separate the term corresponding to $k=0$, integrate in $t$ and expand about $\epsilon=0$ we obtain
\begin{align}
    &U(p,p,\epsilon)=\pi^2\,\Gamma(2\epsilon)\ \sum_{l\in\mathbb{Z}^4}
    \frac{e^{2\pi i\,l\theta p}}{\{(l+p)^2+m^2\}^{1+\epsilon}}
    \int_0^1 du\,\frac{(u(1-u))^\epsilon}{(m^2+u(1-u)l^2)^{2\epsilon}}+\mbox{}\nonumber\\
    &\mbox{}+
    2\pi^2\sum_{l}\frac{e^{2\pi i\,l\theta p}}{(l+p)^2+m^2}\sum_{k\neq 0}\int_0^1 du
    \,e^{-2\pi i u lk}
    \,K_0(2\pi |k|\sqrt{m^2+u(1-u)l^2})+O(\epsilon)
    \,.
\end{align}
Note that for $p\notin\mathcal{Z}_\theta$ the only nonlocal divergence in this expression is contained in the first term, so that
\begin{align}
    U(p,p,\epsilon)&=\frac{\pi^2}{2\epsilon}\,S_1(p)+{\rm quad.\, pol.}+O(\epsilon)\,.
\end{align}
In particular, for irrational $\theta$ this expression holds for any $p\neq 0$ and its divergent part is cancelled by \eqref{lambda2l}; the contribution $U(0,0)$ is, of course, regularized by the renormalization of the parameter $\mu^2$.

\section{The double sum $T(p)$}\label{ap_tp}

The double sum
\begin{align}
        T(p,\epsilon)=\sum_{k,l\in\mathbb{Z}^4}
        \frac{e^{2\pi i\,k\theta l}}{\{[(k+p)^2+m^2](l^2+m^2)^2\}^{1+\epsilon}}
\end{align}
can be written as
\begin{align}
            T(p,\epsilon)=\frac 1{(p^2+m^2)^{1+\epsilon}}\,S_2(0,\epsilon)+
            \sum_{k\neq 0}\sum_l\frac{e^{2\pi i\,k\theta l}}{\{[(k+p)^2+m^2](l^2+m^2)^2\}^{1+\epsilon}}\,.
\end{align}
We will show that the divergent part of the second term in this expression does not depend on $p$ or, equivalently, that the expression
\begin{align}\label{ETp}
        \sum_{k\neq 0}\sum_l\frac{e^{2\pi i\,k\theta l}}{(l^2+m^2)^{2(1+\epsilon)}}
        \left(\frac{1}{[(k+p)^2+m^2]^{1+\epsilon}}-\frac{1}{(k^2+m^2)^{1+\epsilon}}\right)
\end{align}
is finite at $\epsilon=0$. In order to do that, we expand \eqref{ETp} in Taylor series in $p$ keeping potentially divergent terms only,
\begin{align}
        \sum_{k\neq 0}\sum_l \frac{e^{2\pi i k\theta l}}{(l^2+m^2)^{2(1+\epsilon )}}
        \left( -\frac{(1+\epsilon)\, p^2}{(k^2+m^2)^{2+\epsilon}}
        +\frac{2(1+\epsilon)(2+\epsilon)\,(k\cdot p)^2}{(k^2+m^2)^{3+\epsilon}}\right)\,. \label{ETp2}
\end{align}
Assuming $\theta$ is proportional to the standard four-dimensional symplectic matrix times a constant\footnote{In other words, we assume that the NC torus has two sets of noncommuting coordinates with the same noncommutativity parameter.} we can replace $(k\cdot p)^2$ by $\tfrac 14 p^2k^2$, which makes \eqref{ETp2} convergent.

Finally, we show that the difference between expressions \eqref{ETp} and \eqref{ETp2} is finite at $\epsilon=0$. In fact, after Poisson inversion this difference reads
\begin{align}
        2\pi^2 \sum_{k\neq 0}\sum_l K_0(2\pi |l+\theta k|)
        \left[ \frac 1{(k+p)^2+m^2} -\frac 1{k^2+m^2}
        +\frac{p^2}{(k^2+m^2)^2}-\frac{4 (k\cdot p)^2}{(k^2+m^2)^3} \right]\,, \label{ETp3}
\end{align}
where the expression in square brackets is $O(|k|^{-6})$ for large $|k|$. Now, for any given $k$ one can separate in the $l$-sum the term corresponding to the smallest value of $|l+\theta k|$. Under the Diophantine condition, this term decreases as $|k|^{-6}\,\log{|k|}$, whereas the remaining terms are bounded by $e^{-2\pi |l+\theta k|}$, due to the behavior of $K_0$ for large arguments. The subsequent sum in $k$ is therefore absolutely convergent.

\end{document}